\documentclass{aastex63}

\newcommand{\Alfven}{Alfv\'en }
\newcommand{\Alfvenp}{Alfv\'en}
\graphicspath{{./}{figures/}}

\received{\today}
\revised{???}
\accepted{???}
\submitjournal{ApJ}

\shorttitle{Wind from WD J005311}
\shortauthors{Kashiyama et al.}

\begin{document}

\title{The optically thick rotating magnetic wind from a massive white dwarf merger product}

\correspondingauthor{Kazumi Kashiyama}
\email{kashiyama@phys.s.u-tokyo.ac.jp}

\author[0000-0003-4299-8799]{Kazumi Kashiyama}
\affil{Research Center for the Early Universe, Graduate School of Science, University of Tokyo, Bunkyo-ku, Tokyo 113-0033, Japan}
\affil{Department of Physics, Graduate School of Science, University of Tokyo, Bunkyo-ku, Tokyo 113-0033, Japan}

\author{Kotaro Fujisawa}
\affil{Research Center for the Early Universe, Graduate School of Science, University of Tokyo, Bunkyo-ku, Tokyo 113-0033, Japan}

\author{Toshikazu Shigeyama}
\affil{Research Center for the Early Universe, Graduate School of Science, University of Tokyo, Bunkyo-ku, Tokyo 113-0033, Japan}



\begin{abstract}
WD J005311 is a newly identified white dwarf (WD) in a mid-infrared nebula. 
The spectroscopic observation indicates the existence of a neon-enriched carbon/oxygen wind with a terminal velocity of $v_{\infty,\rm obs}\sim 16,000\,\rm km\,s^{-1}$ and a mass loss rate of $\dot M_{\rm obs}\sim 3.5\times 10^{-6}\,M_\odot$ yr$^{-1}$.
Here we consistently explain the properties of WD J005311 using a newly constructed wind solution, 
where the optically thick outflow is launched from the carbon burning shell on an oxygen-neon core and accelerated by the rotating magnetic field to become supersonic and unbound well below the photosphere.  
Our model implies that WD J005311 has a mass of $M_* \sim 1.1\mbox{-}1.3\,M_\odot$, a magnetic field of $B_* \sim (2\mbox{-}5)\times 10^7\,\rm G$, and a spin angular frequency of  $\Omega \sim 0.2\mbox{-}0.5 \,\rm s^{-1}$. 
The large magnetic field and fast spin support the carbon-oxygen WD merger origin.  
WD J005311 will neither explode as a type Ia supernova nor collapse into a neutron star. 
If the wind continues to blow another few kyr, WD J005311 will spin down significantly and join to the known sequence of slowly-rotating magnetic WDs.
Otherwise it may appear as a fast-spinning magnetic WD and could be a new high energy source. 
\end{abstract}



\section{Introduction}\label{sec:intro}
Merger products of white dwarfs (WDs) composed of carbon and oxygen have been thought to result in type Ia supernovae~\citep[e.g.,][]{1984ApJ...277..355W,1984ApJS...54..335I}, neutron stars~\citep[e.g.,][]{1985A&A...150L..21S,2012ApJ...748...35S}, rejuvenated hydrogen-deficient stars \citep{1989ApJ...342..430I}, or strongly magnetized massive WDs~\citep[e.g.,][]{1988PASP..100.1302L,1997MNRAS.292..205F}. Though all of these consequences are of particular importance in astrophysics, we do not reach consensus on what conditions determine the fate partly because of a limited chance to directly observe the merger products. 

Recently \cite{2019Natur.569..684G} reported observation of an extremely hot WD in a mid-infrared nebula J005311 that exhibits some features predicted for a merger product. 
In addition, the spectroscopic analysis indicates the existence of a peculiar wind, which might give a clue to the fate of this object.  
In this paper, we call this object WD J005311. 
The characteristics can be summarized as follows. 
\begin{itemize}
\item The effective temperature of WD J005311 is $T_{\rm eff} = 211,000^{+40,000}_{-23,000}\,\rm K$.
\item The distance to WD J005311 is $d = 3.07^{+0.34}_{-0.28}\,\rm kpc$ and the bolometric luminosity is calculated as $\log(L_{\rm bol}/L_\odot) = 4.60\pm0.14$. 
\item The photospheric radius is estimated to be $r_{\rm ph, obs} = 0.15\pm0.04\,R_\odot$.
\item The chemical composition of the wind is dominated by carbon and oxygen ($X_{\rm C} = 0.2\pm0.1$ and $X_{\rm O} = 0.8\pm0.1$).
 \item The neon mass fraction ($X_{\rm Ne} = 0.01$) can be significantly larger than the solar abundance while the iron group mass fraction $(X_{\rm Fe} = 1.6\times 10^{-3})$  is consistent with that. 
\item From the width and strength of the O\,{\footnotesize VI} emission lines, the terminal velocity and mass-loss rate are estimated as $v_{\infty, \rm obs} =16,000\pm1,000\,\rm km\,s^{-1}$ and $\dot M_{\rm obs} = (3.5\pm0.6)\times 10^{-6}\,M_\odot\,{\rm yr^{-1}}$, respectively. Note that $v_{\infty, \rm obs}$ is significantly larger than the escape velocity at the photospheric radius, $v_{\rm esc}(r_{\rm ph, obs})\sim 1,600$ km s$^{-1}$.
\item The apparent size of the infrared nebula is $\Delta \theta \sim 75\,\rm arcsec$, which corresponds to $r_{\rm nb} \sim 1.6\,\rm pc$. From the expansion velocity of $v_{\rm nb} \sim 100\,\rm km\,s^{-1}$, the age of the nebula is estimated to be $\sim 16,000\,\rm yr$. 
\end{itemize}
\cite{2019Natur.569..684G} proposed that WD J005311 is a super-Chandrasekhar-mass remnant of double degenerate carbon-oxygen (CO) WD merger
from the fact that the observed $T_{\rm eff}$ and $L_{\rm bol}$ and the absence of hydrogen and helium broadly consistent with the stellar evolution calculation of such system~\citep{2016MNRAS.463.3461S}. 
Since the merger remnant can have a large magnetic field and fast spin~\citep[e.g.,][]{2013ApJ...773..136J}, the extremely fast wind can be powered by the rotating magnetic field; 
instead of the radiation pressure gradient, the wind is mainly accelerated by the magnetic torque and pressure gradient. 
\cite{2019Natur.569..684G} further proposed that WD J005311 will collapse into a neutron star within the next few $\rm kyr$.
To test this interesting possibility, it is important to clarify the physical properties of the wind and investigate the consequence of the mass loss.

\cite{2019Natur.569..684G} 
assume that WD J005311 has the same stellar structure as \cite{2016MNRAS.463.3461S}, 
consisting of a degenerate core and a gravitationally bound convective envelope, 
and that the base of the wind is the surface or the photosphere, $r_{\rm ph, obs} = 0.15\,R_\odot$. 
They argue that the terminal velocity of the wind can be explained by the rotating magnetic wind model with a surface magnetic field of $B_* \sim 10^8\,\rm G$ 
and the wind is mainly accelerated up to the \Alfven radius $r_{\rm A} \sim 10 \times r_{\rm ph, obs} = 1.5\,R_\odot$.
However, they have not constructed a stellar model consistently from the degenerate core to the wind region.

The chemical abundance of the WD J005311 wind is broadly consistent with so-called neon novae, 
in which some matter is thought to be dredged up from the core of the underlying oxygen-neon (ONe) WD to enrich the envelope with neon ~\citep{1986ApJ...308..721T, 2016ApJ...816...26H}.
Similar situation can be realized in the merger remnant of a CO WD binary~\citep{2016MNRAS.463.3461S};
in the merged CO WD, carbon is ignited off-center and the carbon-burning flame propagates into the interior. 
The flame reaches the center in $\sim$ 10 kyr after the merger. 
Then, neutrino cooling leads to the Kelvin-Helmholtz contraction of the ONe core and a series of off-center carbon flashes occur. 
Note that the timing is consistent with the nebula age of J005311.

\begin{figure}\label{fig:cartoon}
\includegraphics[width = \textwidth]{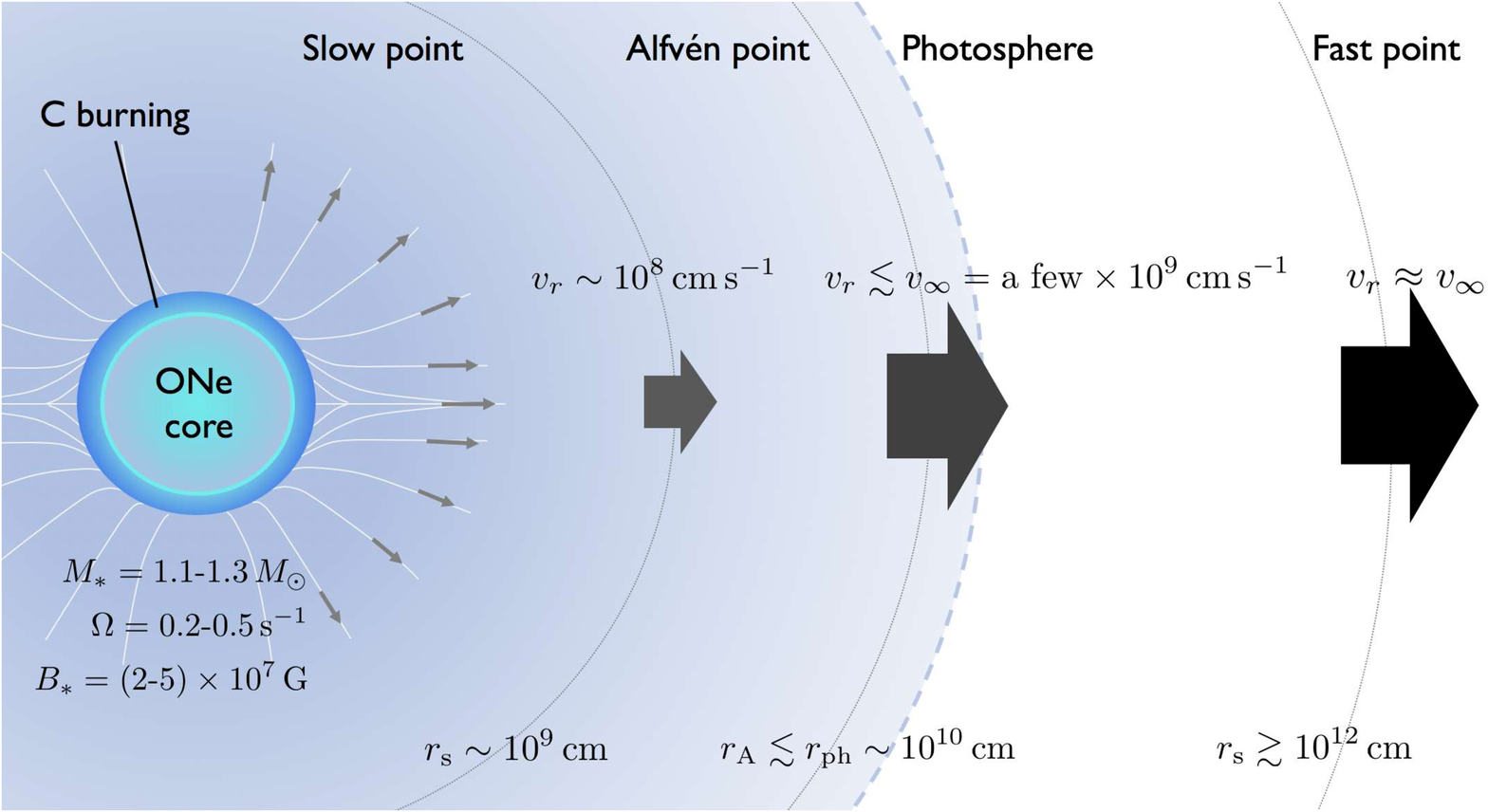}
\caption{
Schematic picture of our wind model for white dwarf J005311. 
}
\end{figure}

Motivated by the chemical abundance, here we synthesize an alternative wind model for WD J005311 
by solving radiation magnetohydrodynamic equations consistently from the degenerate core surface to well beyond the \Alfven radius.
Fig. \ref{fig:cartoon} shows the schematic picture.  
We regard these carbon flashes are responsible not only for neon rich materials in the envelope but for the launch of the wind. 
After the launch, the wind is mainly accelerated by the rotating magnetic field and becomes supersonic and unbound below the photosphere ($r_{\rm A} \lesssim 0.15\,R_\odot$).
In the following section, we build the wind equation compatible with our scenario, 
and solve the equation to find solutions that can explain the observed properties of WD J005311. 


\section{The rotating magnetic wind model for WD J005311} \label{sec:model}
 We show the basic equations of our wind model in Sec. \ref{sec:eqs} and consider the boundary conditions in Sec. \ref{sec:bc}. 
 After briefly describing the calculation method in Sec. \ref{sec:method}, we show the wind solutions for WD J005311 in Sec. \ref{sec:results}.
 
\subsection{Basic Equations}\label{sec:eqs}
We basically combine the classical Weber-Davis stellar wind solution~\citep{1967ApJ...148..217W} and the optically thick nova wind model~\citep{1994ApJ...437..802K} both of which assume stationary state.
As the Weber-Davis solution, we only consider the wind in the equatorial plain emanating from the surface of a rigidly rotating star with a mass $M_*$ and an angular frequency $\Omega$. The flux freezing condition together with Faraday's law gives the flux conservation of the radial magnetic field $B_r$ and a relation with the toroidal magnetic field $B_\phi$ as 
\begin{equation}\label{eq:Br_con}
    {\cal F}_B = r^2 B_r = {\rm const.},
\end{equation}
and 
\begin{equation}
    \frac{B_\phi}{B_r} = \frac{v_\phi - r\Omega}{v_r}.
\end{equation}
Here $v_r$ and $v_\phi$ are the radial and azimuthal velocity of the wind, respectively, and $r$ denotes the radial coordinate. 
The mass and angular momentum conservations are described as 
\begin{equation}\label{eq:mass_con}
    \frac{\dot M}{4\pi} = \rho v_r r^2 = {\rm const.},
\end{equation}
\begin{equation}\label{eq:L_con}
    {\cal L} = r v_\phi - \left(\frac{rB_rB_\phi}{4\pi \rho v_r}\right) = {\rm const.},
\end{equation}
by introducing the mass loss rate $\dot M$ and the specific angular momentum $\cal L$, where $\rho$ is the mass density of the wind. 
The coupling between the gas and radiation is treated by means of  the flux-limited diffusion approximation, 
where the temperature gradient can be described as
\begin{equation}\label{eq:dTdr}
    \frac{dT}{dr} = -\frac{\kappa \rho L_{\rm rad}}{16\pi a c \lambda T^3 r^2},
\end{equation}
where $\kappa$ is the opacity, $L_{\rm rad}$ the radiation luminosity, $a$ the radiation constant, $c$ the speed of light, $T$ the temperature, and $\lambda$ is the flux limiter. 
We calculate $\kappa = \kappa(\rho, T)$ using the OPAL code~\citep{1996ApJ...464..943I}, 
assuming the metal abundance consistent with the J005311 wind,  
and $\lambda = \lambda(\rho,\kappa,d\ln T/dr)$ following \cite{1981ApJ...248..321L}. 
The momentum equation of the gas in the radial direction is given by
\begin{equation}\label{eq:eom_r}
    v_r \frac{dv_r}{dr}+\frac{1}{\rho}\frac{dP_{\rm gas}}{dr} - \frac{\kappa L_{\rm rad}}{4\pi r^2 c}  + \frac{GM_*}{r^2} -\frac{v_\phi{}^2}{r} + \frac{B_\phi}{4\pi \rho r}\frac{d}{dr}(rB_\phi) = 0.
\end{equation}
The first four terms are identical to the spherical radiation driven wind; 
the first, second, third, and fourth terms correspond to gas advection, gas pressure gradient, radiation pressure gradient, and gravitational acceleration, respectively. 
The gas pressure is given by $P_{\rm gas} = \rho k_{\rm B}T/(\mu m_{\rm u})$ where $\mu$ is the mean molecular weight, $k_{\rm B}$ the Boltzmann constant, and $m_{\rm u}$ is the atomic mass unit. 
The fifth and sixth terms in Eq. (\ref{eq:eom_r}) are the radial acceleration produced by the centrifugal force and the magnetic sling effect, respectively. 
The energy conservation equation is given by 
\begin{equation}\label{eq:ene_con}
v_r \frac{d\varepsilon_{\rm gas}}{dr} + P_{\rm gas} v_r \frac{d}{dr} \left(\frac{1}{\rho}\right) = - \frac{1}{4\pi r^2 \rho} \frac{dL_{\rm rad}}{dr},
\end{equation}
where $\varepsilon_{\rm gas} = 3k_{\rm B}T/(2\mu m_{\rm u})$ is the gas energy density.
Note that we neglect the convective luminosity in Eq (\ref{eq:ene_con}). 
In general, there exists a convective interface between the carbon burning region and the wind region. 
However, the convective luminosity is relevant only in the vicinity of the surface region as in the case of nova winds with a similar mass loss rate~\citep{1994ApJ...437..802K}. 
Also in our case, the convective motion might be suppressed by strong magnetic fields~\citep[e.g.,][]{2014Natur.515...88V}.

Using Eqs. (\ref{eq:Br_con}-\ref{eq:L_con}),  Eqs. (\ref{eq:eom_r}) and (\ref{eq:ene_con}) can be transformed into 
\begin{equation}\label{eq:eom_r'}
   \left(v_r^2 - \frac{k_{\rm B}T}{\mu m_{\rm u}} - \frac{A_\phi^2 v_r^2}{v_r^2 -A_r^2}\right) \frac{r}{v_r} \frac{dv_r}{dr} 
   = \frac{\kappa L_{\rm rad}}{4\pi r c} + \frac{k_{\rm B}}{\mu m_{\rm u}}\left(\frac{dT}{d\log r} + 2T\right) - \frac{GM_*}{r}  +  v_\phi^2 + 2 v_r v_\phi \frac{A_r A_\phi}{v_r^2 - A_r^2}, 
\end{equation}
and 
\begin{equation}\label{eq:ene_con'}
\frac{d\varepsilon}{dr} = \frac{\kappa L_{\rm rad}}{4\pi r^2 c}, 
\end{equation}
with 
\begin{equation}\label{eq:Ar}
A_r = \frac{B_r}{\sqrt{4\pi \rho}},
\end{equation}
being the radial \Alfven velocity,
\begin{equation}
A_\phi = \frac{B_\phi}{\sqrt{4\pi \rho}},
\end{equation}
being the azimuthal \Alfven velocity, and 
\begin{equation}\label{eq:e_eff}
\varepsilon = \frac{L_{\rm rad}}{\dot M} + \frac{1}{2}(v_r{}^2+v_\phi{}^2)  + \frac{5}{2}\frac{kT}{\mu m_u} - \frac{GM_*}{r}  - r\Omega v_\phi +{\cal L}\Omega,
\end{equation}
respectively. 
In total, we have 4 constraint equations (Eqs. \ref{eq:Br_con}-\ref{eq:L_con}) and 3 ordinary differential equations (Eqs. \ref{eq:dTdr}, \ref{eq:eom_r'}, and \ref{eq:ene_con'}) for 7 variables $(\rho,v_r,v_\phi,B_r,B_\phi,T,L_{\rm rad})$.

\subsection{Boundary conditions}\label{sec:bc}
A rotating magnetic wind solution has to satisfy the regularity condition at the slow, \Alfvenp, and fast points~\citep[e.g.,][]{1999isw..book.....L}.
It is convenient to normalize variables with the quantities at the \Alfven radius $r_{\rm A}$ where the \Alfven Mach number becomes unity, i.e., $v_r = A_r$. 
From Eqs (\ref{eq:Br_con}), (\ref{eq:mass_con}) and (\ref{eq:Ar}), the velocity can be written as  
\begin{equation}\label{eq:vA}
v_{\rm A} = A_r (r_{\rm A}) = \frac{{\cal F}_B^2}{\dot M r_{\rm A}^2}.
\end{equation}
Normalizing $r$ and $v_r$ with $r_{\rm A}$ and $v_{\rm A}$ as $x \equiv r/r_{\rm A}$ and $u \equiv v_r/v_{\rm A}$,
$\rho$, $B_r$, $v_{\phi}$ and $B_{\phi}$ can be described as 
\begin{equation}\label{eq:rho}
\rho = \frac{\dot M}{4 \pi v_{\rm A} r_{\rm A}^2} \frac{1}{ux^2}
\end{equation}
\begin{equation}
B_r = \frac{{\cal F}_B}{r_{\rm A}^2} \frac{1}{x^2},
\end{equation}
\begin{equation}
v_{\phi} = r_{\rm A}\Omega \frac{x(1-u)}{1-x^2u},
\end{equation}
\begin{equation}\label{eq:Bphi}
B_{\phi} = - \frac{{\cal F}_B \Omega}{r_{\rm A} v_{\rm A}} \frac{1-x^2}{x(1-x^2u)}.
\end{equation}
We note that the angular momentum can be also simplified as ${\cal L} = r_{\rm A}^2 \Omega$.
The regularity of $v_\phi$ and $B_\phi$ and also Eq. (\ref{eq:eom_r'}) at $r = r_{\rm A}$ can be guaranteed by introducing the slope $du/dx|_{r = r_{\rm A}}$ as an additional parameter.
In the limit of $x \rightarrow 1$ and $u \rightarrow 1$, $v_\phi$ and $B_\phi$ become 
\begin{equation}
v_\phi (r_{\rm A}) = r_{\rm A}\Omega \frac{du/dx|_{r = r_{\rm A}}}{2+du/dx|_{r = r_{\rm A}}},
\end{equation}
\begin{equation}
B_\phi (r_{\rm A}) =  - \frac{{\cal F}_B \Omega}{r_{\rm A} v_{\rm A}} \frac{2}{2+du/dx|_{r = r_{\rm A}}}. 
\end{equation}
By solving Eqs. (\ref{eq:dTdr}), (\ref{eq:eom_r'}), (\ref{eq:ene_con'}), and (\ref{eq:rho}-\ref{eq:Bphi}) with fixing ($r_{\rm A}, du/dx|_{r = r_{\rm A}}, \dot M, T(r_{\rm A}), L_{\rm rad}(r_{\rm A}), {\cal F}_B, \Omega, M_*$), 
we can obtain one numerical solution. 
Most of the numerical solutions passing $r = r_{\rm A}$ are not physical wind solution; 
a wind solution also has to pass both the slow and fast points where the coefficient of $dv_r/dr$ in Eq. (\ref{eq:eom_r'}) becomes zero, i.e., $\{v_r^2 -k_{\rm B}T/(\mu m_{\rm u})\}(v_r^2 - A_r^2) - A_\phi^2 v_r^2 = 0$. 
For a solution to be regular at these points, the right hand side of Eq. (\ref{eq:eom_r'}) has to be zero there. 
This gives two additional constraints on the wind solution. 
Here we change $r_{\rm A}$ and $du/dx|_{r = r_{\rm A}}$ until finding a wind solution for a given parameter set of ($\dot M, T(r_{\rm A}), L_{\rm rad}(r_{\rm A}), {\cal F}_B, \Omega, M_*$). 

\begin{figure}\label{fig:m-r}
 \centering
\includegraphics[scale = 0.7]{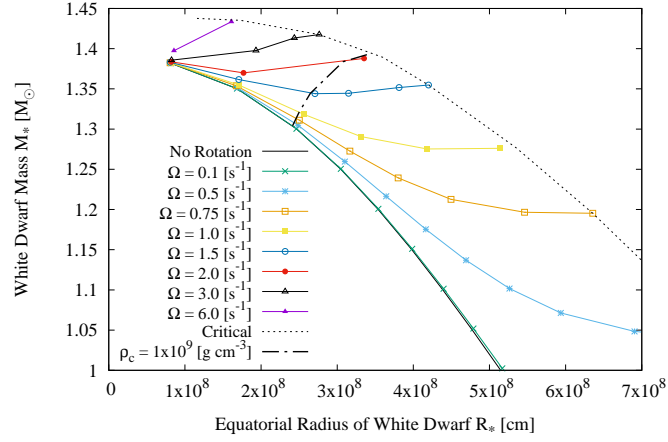}
\caption{
The relation between mass and equatorial radius of oxygen-neon white dwarfs with different angular frequencies.
}
\end{figure}

We have 5 additional constraints to be satisfied by a wind model for WD J005311.
First, the mass loss rate has to be consistent with the observed value; 
\begin{equation}\label{eq:Mdot}
\dot M \gtrsim \dot M_{\rm obs}.
\end{equation}
Second, the terminal velocity of the wind has to be consistent with the observed value; 
\begin{equation}\label{eq:v_r_inf}
v_r (\infty) \gtrsim v_{\infty, \rm obs}.
\end{equation}
Note that we only consider the equatorial direction, for which the mass loss rate and radial velocity are the largest.
As in the case of the Weber-Davis solution, the terminal velocity of our wind solution is approximately given by the so-called Michel velocity~\citep{1969ApJ...158..727M};
\begin{equation}
v_r(\infty) \approx \left(\frac{{\cal F}_B^2 \Omega^2}{\dot M}\right)^{1/3} 
\sim 30,000\,{\rm km\,s^{-1}} \left(\frac{B_*}{10^8\,{\rm G}}\right)^{2/3} \left(\frac{R_*}{1,500\,{\rm km}}\right)^{4/3}  \left(\frac{\Omega}{1\,{\rm s^{-1}}}\right)^{2/3}  \left(\frac{\dot M}{3\times10^{-6}\,M_\odot{\rm yr^{-1}}}\right)^{-1/3},
\end{equation}
where $B_*$ denotes the magnetic field at the surface, i.e., $B_*=B_r(R_*)$. Thus, Eq. (\ref{eq:v_r_inf}) gives an additional constraint on ${\cal F}_B \Omega$. 
Third and fourth, the radiation temperature and luminosity have to be consistent with the observed values; 
\begin{equation}\label{eq:Tcon}
T(r_{\rm ph, obs}) \approx T_{\rm eff},
\end{equation}
\begin{equation}\label{eq:Lr}
L_{\rm rad}(r_{\rm ph, obs}) \approx L_{\rm bol}.
\end{equation}
With using the flux-limited diffusion approximation, there is no perfect definition of the photosphere; the approximation can be only justified for the optically thick and thin limits.
In our wind model for WD J005311, the decoupling between gas and radiation occurs at around the \Alfven radius, where the bulk of the wind acceleration completes.
Thus, instead of Eqs. (\ref{eq:Tcon}) and (\ref{eq:Lr}), we set the conditions on the temperature and luminosity at the \Alfven radius, $T_{\rm A} \equiv T(r_{\rm A})$ and $L_{\rm rad, A} \equiv L_{\rm rad}(r_{\rm A})$ as 
\begin{equation}\label{eq:Tcon}
T_{\rm A} \gtrsim T_{\rm eff},
\end{equation}
\begin{equation}\label{eq:Lr}
L_{\rm rad, A} \approx L_{\rm bol}, 
\end{equation}
with 
\begin{equation}\label{eq:rA}
\left[\frac{L_{\rm rad, A}}{\pi a c T_{\rm A}^4}\right]^{1/2} < r_{\rm A} < r_{\rm ph, \rm obs}. 
\end{equation}
The lower limit of $r_A$ comes from the upper limit of the radiation luminosity, i.e., $L_{\rm rad} < \pi r_{\rm ph, \rm obs}^2 a c T_{\rm eff}^4$.
Fifth, we have an inner boundary condition; at the surface of the degenerate core ($r=R_*$), the wind is powered by the nuclear burning 
and the energy generation rate should be equal to the radiation luminosity; 
\begin{equation}\label{eq:inner_boundary}
L_{\rm nuc}(\rho(R_*),T(R_*),R_*) \approx L_{\rm rad} (R_*).
\end{equation}
For evaluating the left hand side, we use the analytic fitting formula for the energy generation rate by the carbon burning \citep{2012sse..book.....K};
\begin{eqnarray}
\varepsilon_{\rm CC} \approx && 5.49\times 10^{43}\,{\rm erg \ s^{-1} \ g^{-1}}\, f_{\rm CC} \rho X^2_{\rm C}T_9^{-3/2}T_{9\alpha}^{5/6}\exp[-84.165/T_{9\alpha}^{1/3}] \\
&\times& [\exp(-0.01T_{9\alpha}^4)+5.56\times10^{-3}\exp(1.685T_{9\alpha}^{2/3})]^{-1}
\end{eqnarray}
where $T_9 = (T/10^9\,{\rm K})$ and $T_{9\alpha} = T_9/(1+0.0067T_9)$. 
For simplicity, we fix the screening factor as $f_{\rm CC} = 0.5$ and the impacts of neutrino cooling is neglected. 

After putting the conditions given by Eqs. (\ref{eq:Mdot}), (\ref{eq:v_r_inf}), (\ref{eq:Tcon}), (\ref{eq:Lr}), and (\ref{eq:inner_boundary}), the residual model parameters are the mass $M_*$ and (equatorial) radius $R_*$ of the degenerate core. 
When we specify the $M_*$-$R_*$ relation, a sequence of wind solutions for WD J005311 can be obtained. 
To this end, we calculate the 2 dimensional equilibrium structure of rotating ONe WD using the self-consistent field scheme~\citep{2015MNRAS.454.3060F}.
We assume the uniform rotation, and use the zero-temperature equation of state with $Z = 9.5$~\citep{1961ApJ...134..669S} since we are interested in the ONe core after the Kelvin Helmholtz contraction.
Note that the magnetic field can be safely neglected for $B_* < 10^9\,\rm G$~\citep{2012MNRAS.422..434F}.
In this case, an equilibrium structure can be obtained for a given angular frequency and central density. 
The results are shown in Fig. \ref{fig:m-r}.  
Each solid line indicates the $M_*$-$R_*$ relation with each of different angular frequencies ranging from $\Omega = 0\,{\rm s^{-1}}$ to $\Omega = 6\,\rm s^{-1}$. 
The dotted line corresponds to the cases with the critical rotation velocity.  
When the rotation is negligible, i.e., $\Omega \lesssim 0.1\,{\rm s^{-1}}$, $R_*$ ranges from $\sim 1\times 10^8\,{\rm cm}$ to $\sim 5\times 10^{8}\,{\rm cm}$, and the maximum mass is $M_{\rm ch} \sim 1.38\,M_\odot$.
For a given $M_*$, the near critical case has a larger $R_*$ than the non-rotating case by a factor of $\sim 2$. 
For $\Omega \gtrsim 2\,s^{-1}$, the maximum mass of the WD becomes larger than $M_{\rm ch}$ at most by $\lesssim 10 \%$.
As a reference, we show the cases with the central density of $\rho_{\rm c} = 10^9\,\rm g\,cm^{-3}$ with the dotted line.

\subsection{Calculation Methods}\label{sec:method}
The calculation method is as follow. 
\begin{enumerate}
\item First, we fix $\dot M$, ${\cal F}_B \Omega$, $T_{\rm A}$, $L_{\rm rad, A}$ so that Eqs. (\ref{eq:Mdot}\mbox{-}\ref{eq:v_r_inf}) and (\ref{eq:Tcon}\mbox{-}\ref{eq:Lr}) are satisfied.
The observed terminal velocity and mass loss rate should be regarded as the lower limits of $v_r(\infty)$ and $\dot M$. 
Here, we study three cases with $(v_r(\infty), \dot M) = (2.4\times 10^9\,{\rm cm\,s^{-1}},6\times 10^{-6}\,M_\odot\,\rm yr^{-1})$, $(1.9\times 10^9\,{\rm cm\,s^{-1}},6\times 10^{-6}\,M_\odot\,\rm yr^{-1})$, and $(2.4\times 10^9\,{\rm cm\,s^{-1}},3\times 10^{-6}\,M_\odot\,\rm yr^{-1})$.
\item We fix $M_*$ and $\Omega$.
We numerically integrate differential equations (\ref{eq:dTdr}), (\ref{eq:eom_r'}), and (\ref{eq:ene_con'}) with the 4th-order explicit Runge-Kutta method both inward and outward from the \Alfven point. The constraint equations described with dimensionless variables $u$ and $x$ as Eqs. (\ref{eq:rho}-\ref{eq:Bphi}) are consistently solved.
We tune ($r_{\rm A}$, $du/dx|_{r = r_{\rm A}}$) until we obtain a wind solution that passes the slow and fast points.  
\item If the wind solution satisfies the $M_*$-$R_*$ relation, we adapt it as a wind solution for WD J005311. 
If not, we change $M_*$ or $\Omega$ (but still fixing ${\cal F}_B \Omega$) and repeat the above calculations. We explore a range of $1\,M_\odot < M_* < 1.45\,M_\odot$ and $\Omega < 6\,\rm s^{-1}$. 
\end{enumerate}

\subsection{Results}\label{sec:results}
When fixing the observed parameters, $v_{\infty, \rm obs}, \dot M_{\rm obs}, T_{\rm eff}$, and $L_{\rm bol}$, and the $M_*\mbox{-}R_*$ relation, the wind solution for WD J005311 can be uniquely determined, namely $M_*$, $B_*$, and $\Omega$ of this object can be determined.
We search solutions by taking uncertainties in the observed parameters and limitations in our model into account as described in Sec.\ref{sec:method}. 

\begin{figure}
 \begin{minipage}[b]{0.5\linewidth}
  \centering
  \includegraphics[scale = 0.7]{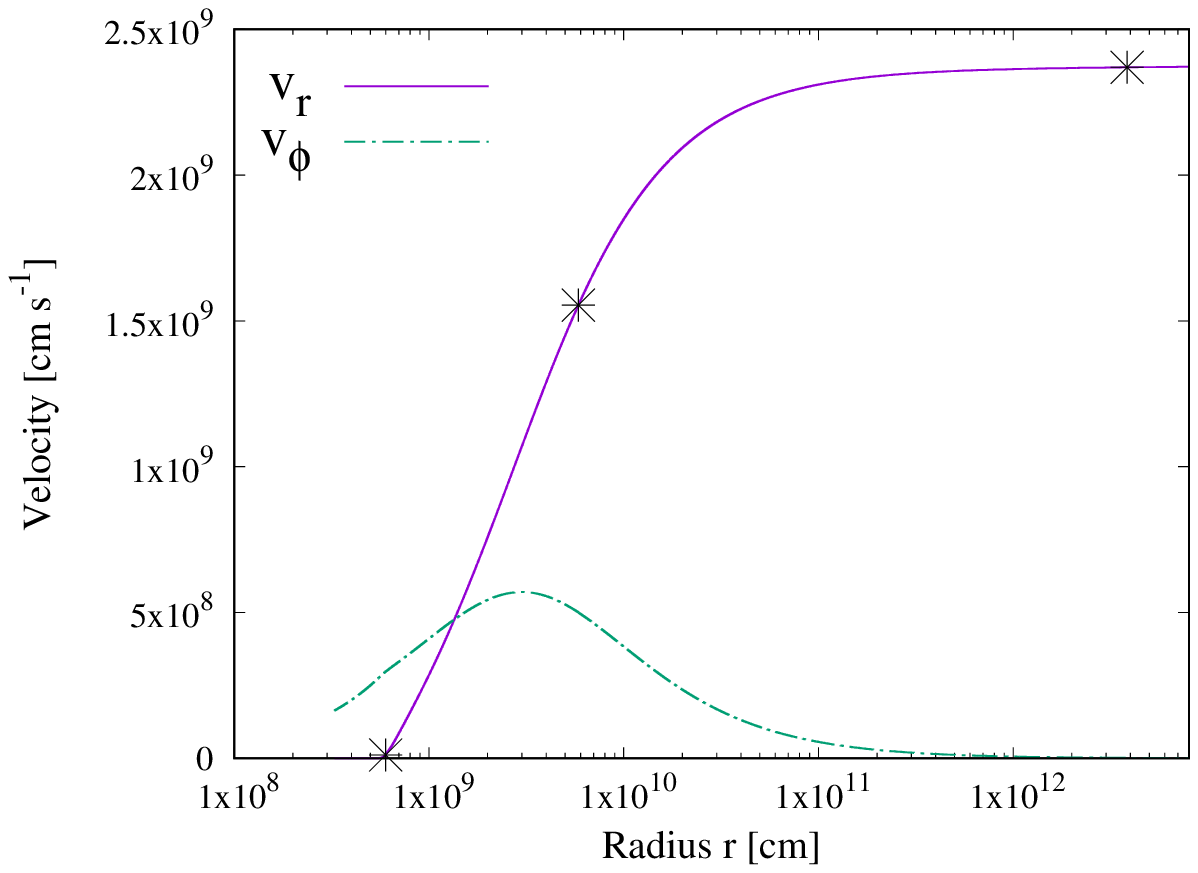}
 \end{minipage}
 \begin{minipage}[b]{0.5\linewidth}
  \centering
  \includegraphics[scale = 0.7]{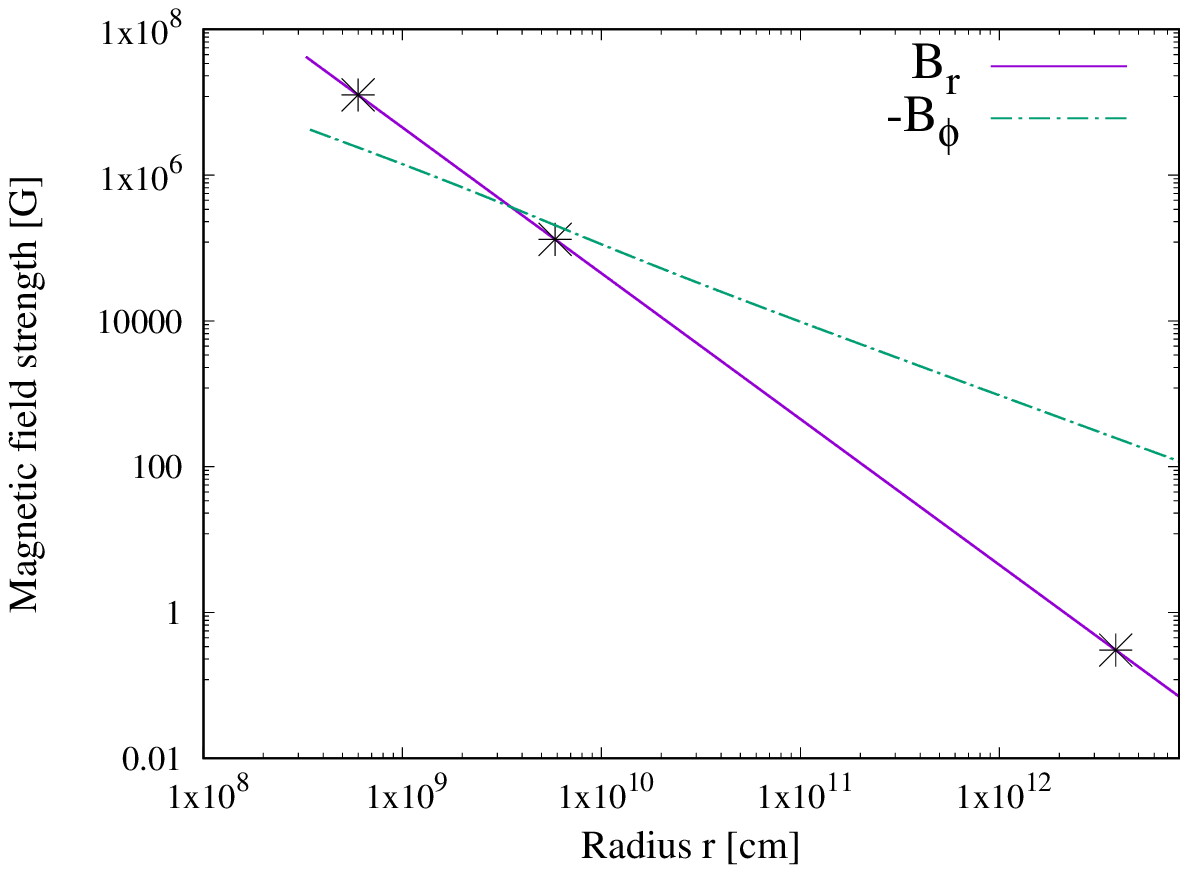}
 \end{minipage}\\
 \begin{minipage}[b]{0.5\linewidth}
  \centering
  \includegraphics[keepaspectratio, scale=0.7]
  {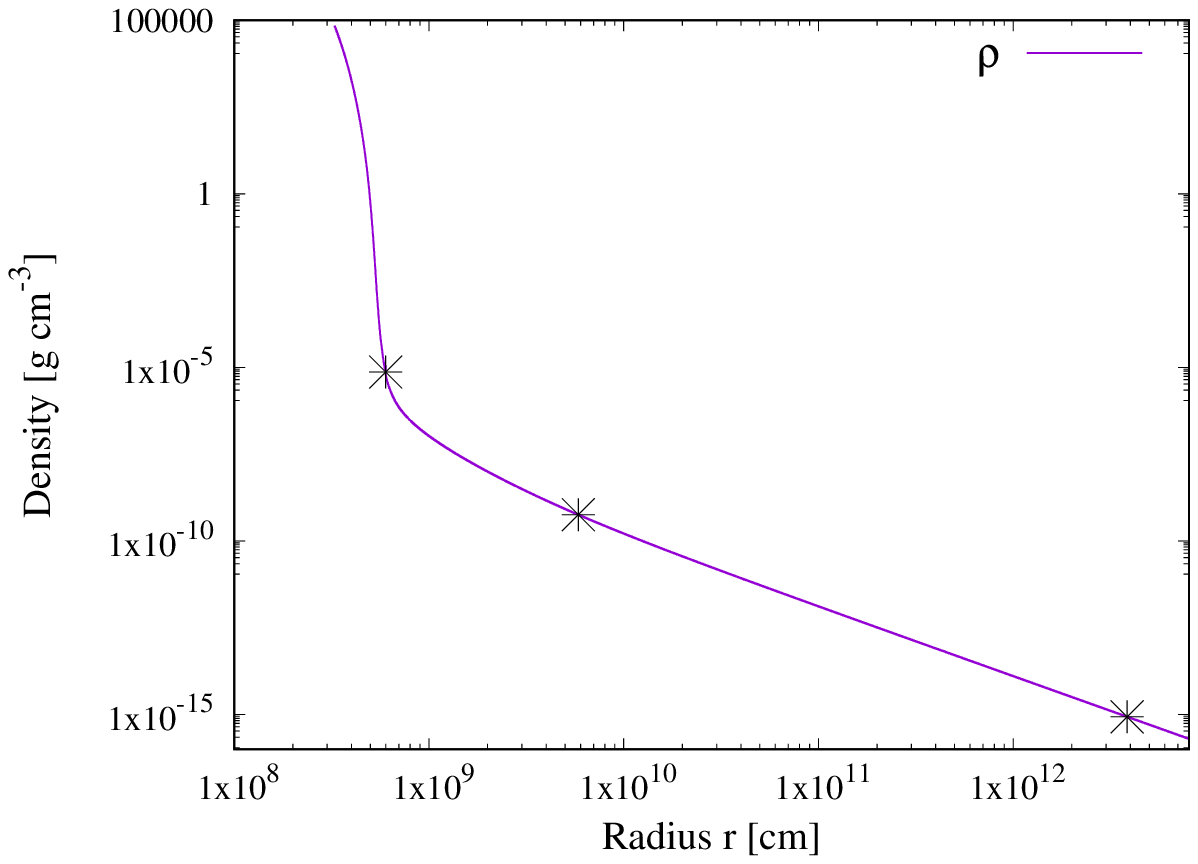}
 \end{minipage}
 \begin{minipage}[b]{0.5\linewidth}
  \centering
  \includegraphics[keepaspectratio, scale=0.7]
  {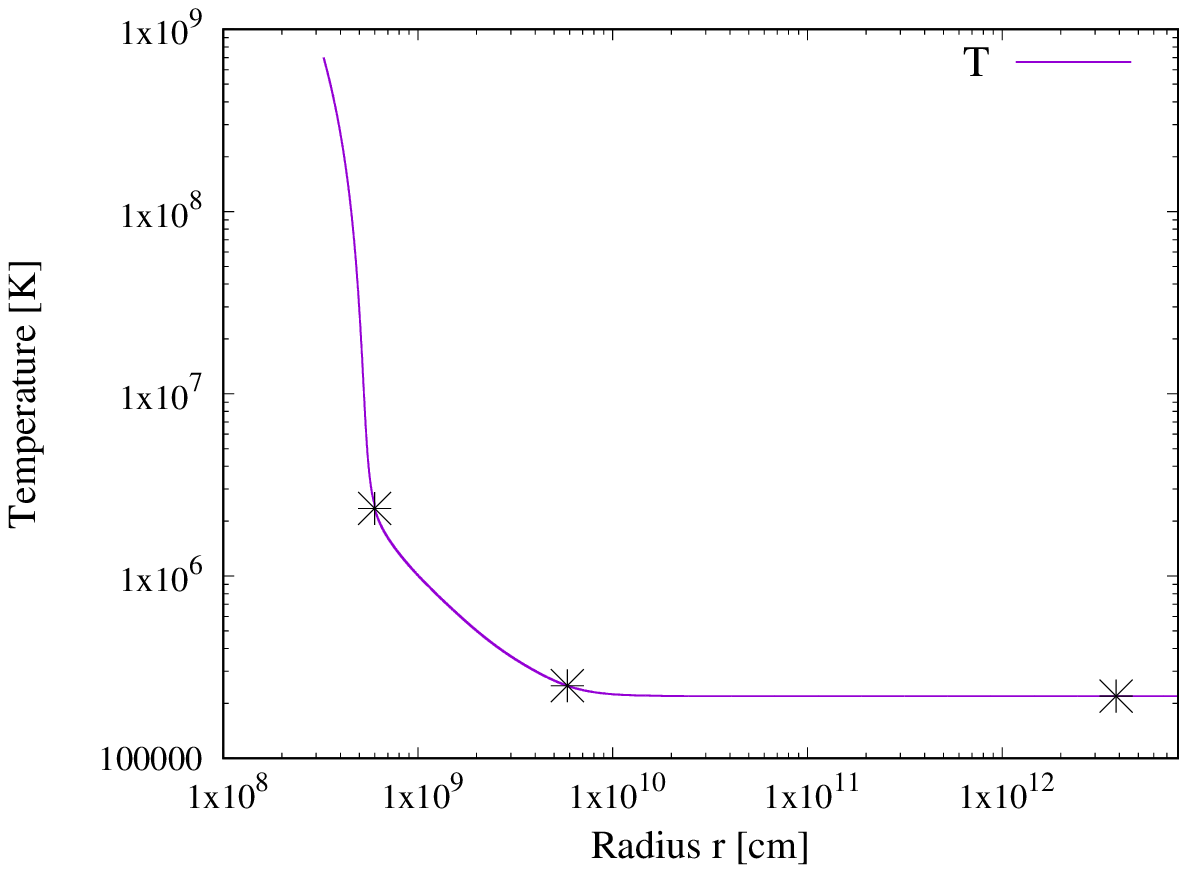}
 \end{minipage}\\
 \begin{minipage}[b]{0.5\linewidth}
  \includegraphics[keepaspectratio, scale=0.7]
  {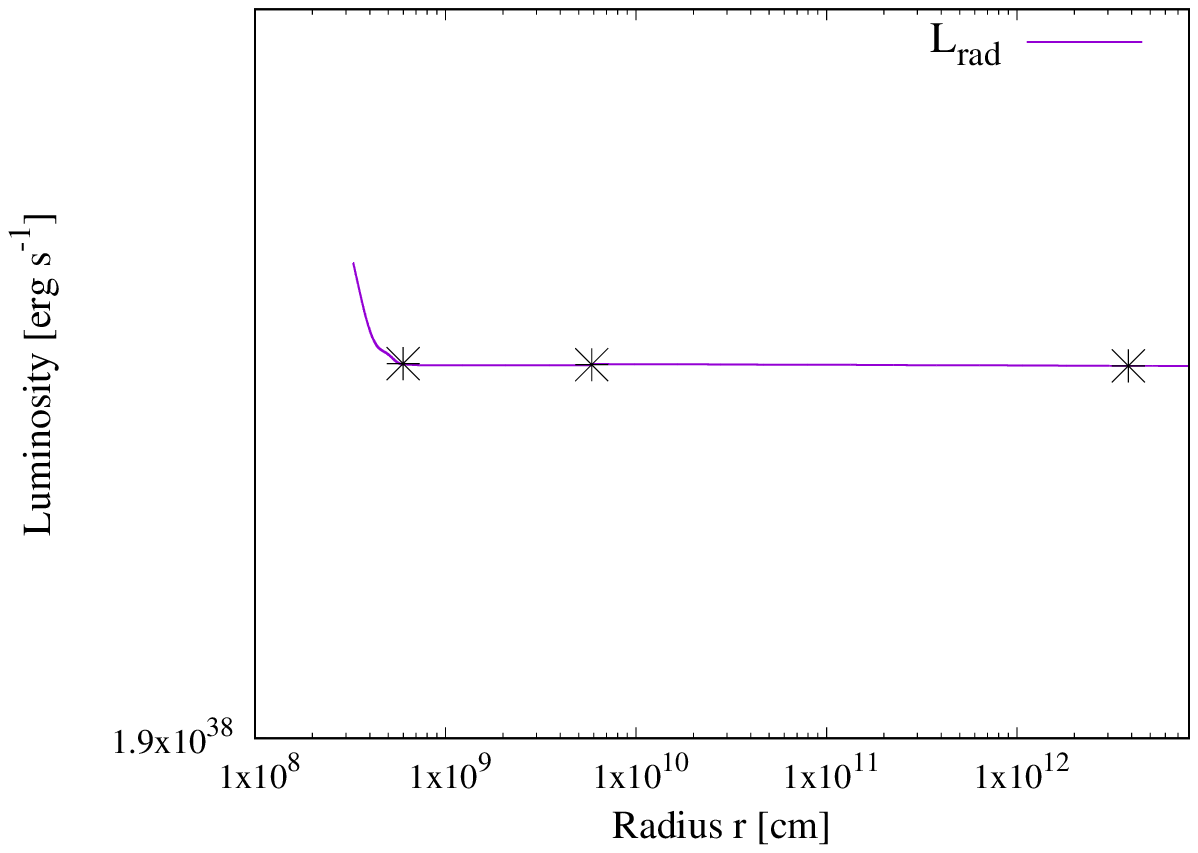}
 \end{minipage}
\caption{
The radial profiles of a rotating magnetic wind solution for WD J005311. We show the radial ($v_r$) and azimuthal ($v_\phi$) velocity, radial ($B_r$) and azimuthal ($B_\phi$) magnetic field strength, gas density ($\rho$), temperature ($T$), and radiation luminosity ($L_{\rm rad}$) for the case with $M_* = 1.25\,M_\odot$, $R_* = 3.3\times 10^{8}\,\rm cm$, $B_* = 4.2 \times 10^7\,$ G, $\Omega = 0.5\,\rm s^{-1}$, and $\dot M = 6\times 10^{-6}\,M_\odot\,\rm yr^{-1}$. The asterisks indicate the positions of the slow, Alfv\'en, and fast points. 
}\label{fig:profile}
\end{figure}

\begin{figure}
\begin{minipage}[b]{0.5\linewidth}
  \centering
  \includegraphics[keepaspectratio, scale=0.7]
  {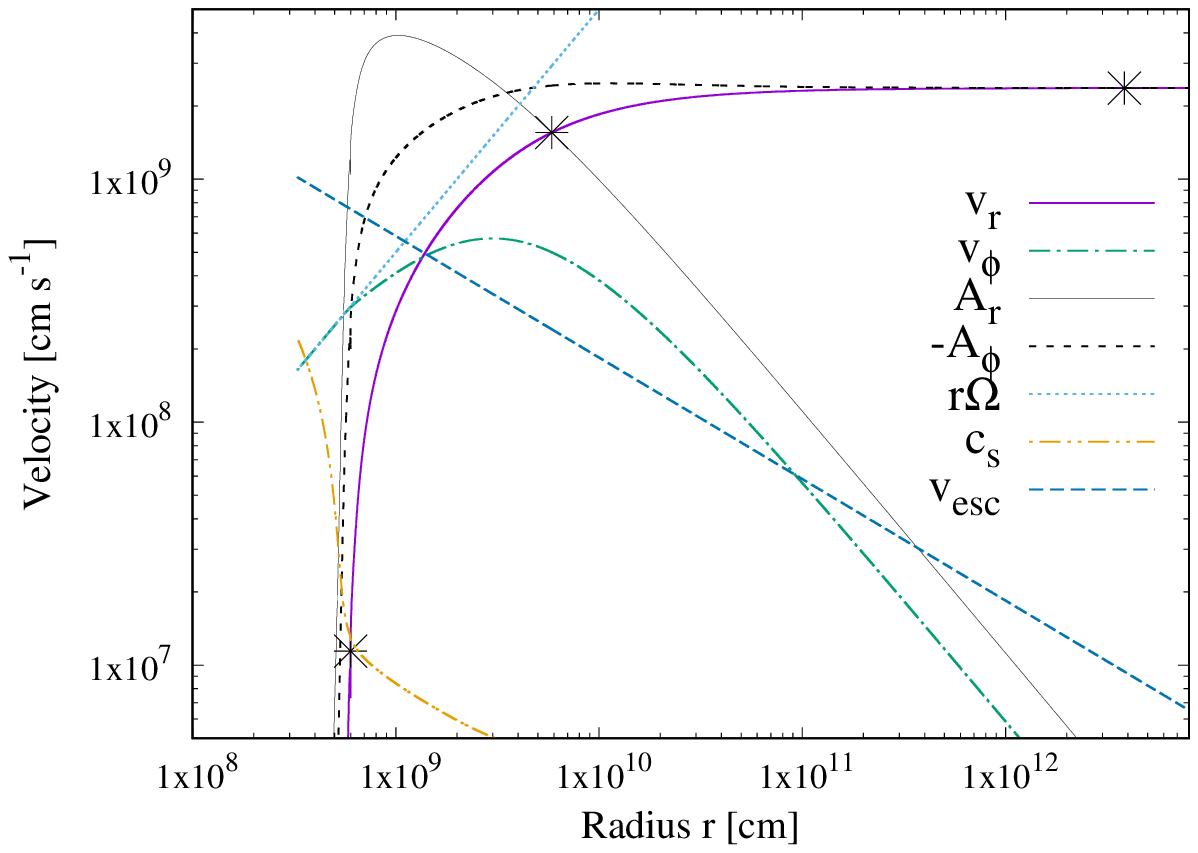}
 \end{minipage}
 \begin{minipage}[b]{0.5\linewidth}
  \centering
  \includegraphics[keepaspectratio, scale=0.7]
  {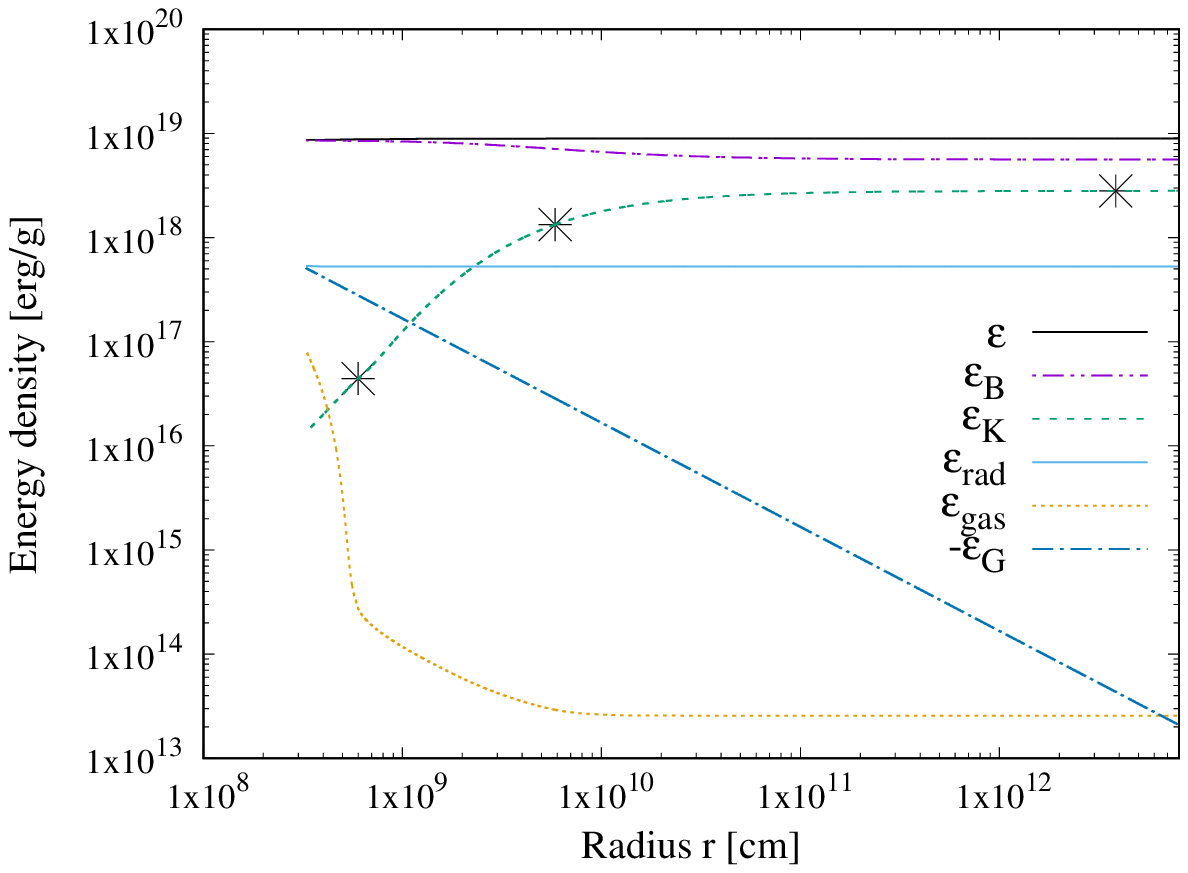}
 \end{minipage}
 \caption{
 {\it Left}: Comparison of the radial and azimuthal velocities of the wind solution in Fig. \ref{fig:profile} with other relevant velocities including the radial ($A_r$) and azimuthal ($A_\phi$) \Alfven velocities, the corotation velocity with the degenerate core ($r\Omega$), the adiabatic sound velocity of the gas ($c_{\rm s}$), and the escape velocity ($v_{\rm esc}$).
 {\it Right}: The radial profile of the energy density of the wind solution in Fig. \ref{fig:profile}. We show the contributions from magnetic field ($\varepsilon_B$), kinetic energy ($\varepsilon_{\rm K}$), radiation ($\varepsilon_{\rm rad}$), thermal gas ($\varepsilon_{\rm gas}$), and gravitational energy ($\varepsilon_{G}$).
 }\label{fig:comp}
\end{figure}

\begin{figure}
 \begin{minipage}[b]{0.5\linewidth}
  \centering
  \includegraphics[keepaspectratio, scale=0.7]
  {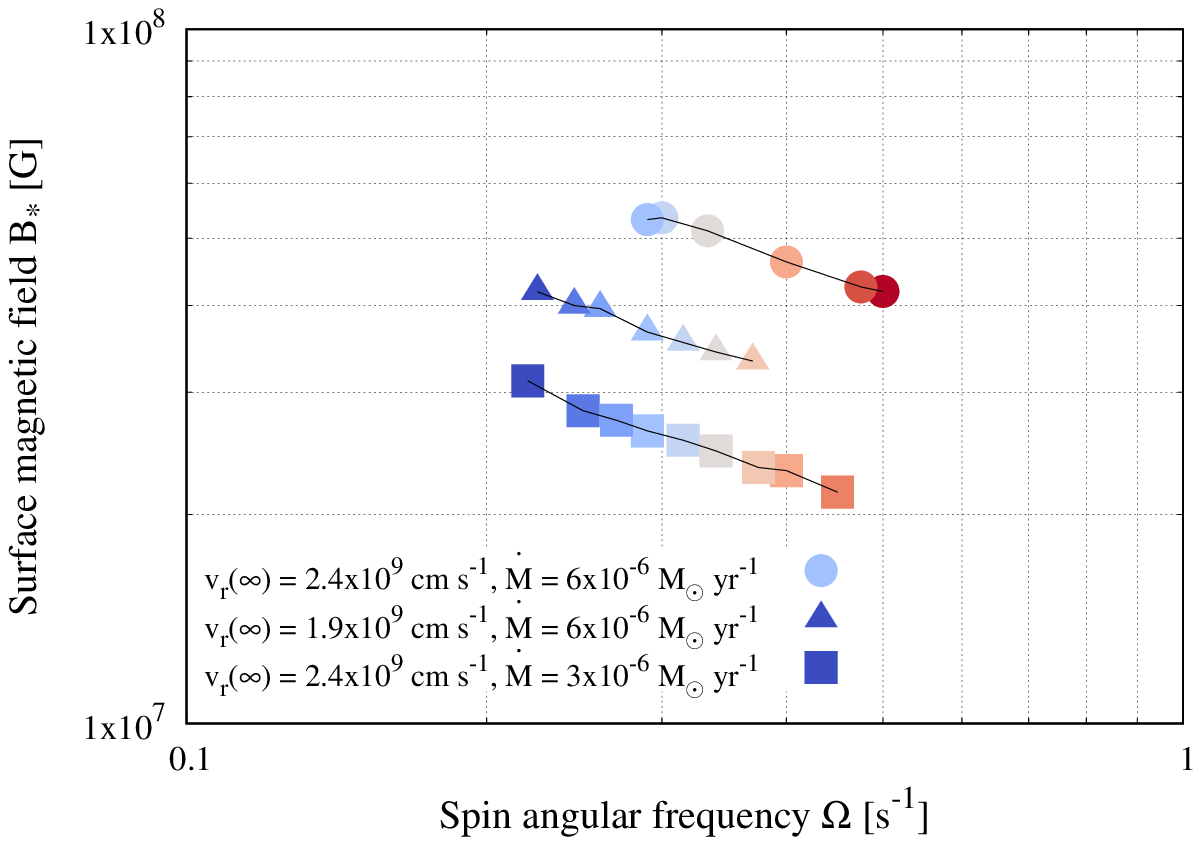}
 \end{minipage}
 \begin{minipage}[b]{0.5\linewidth}
  \centering
  \includegraphics[keepaspectratio, scale=0.7]
  {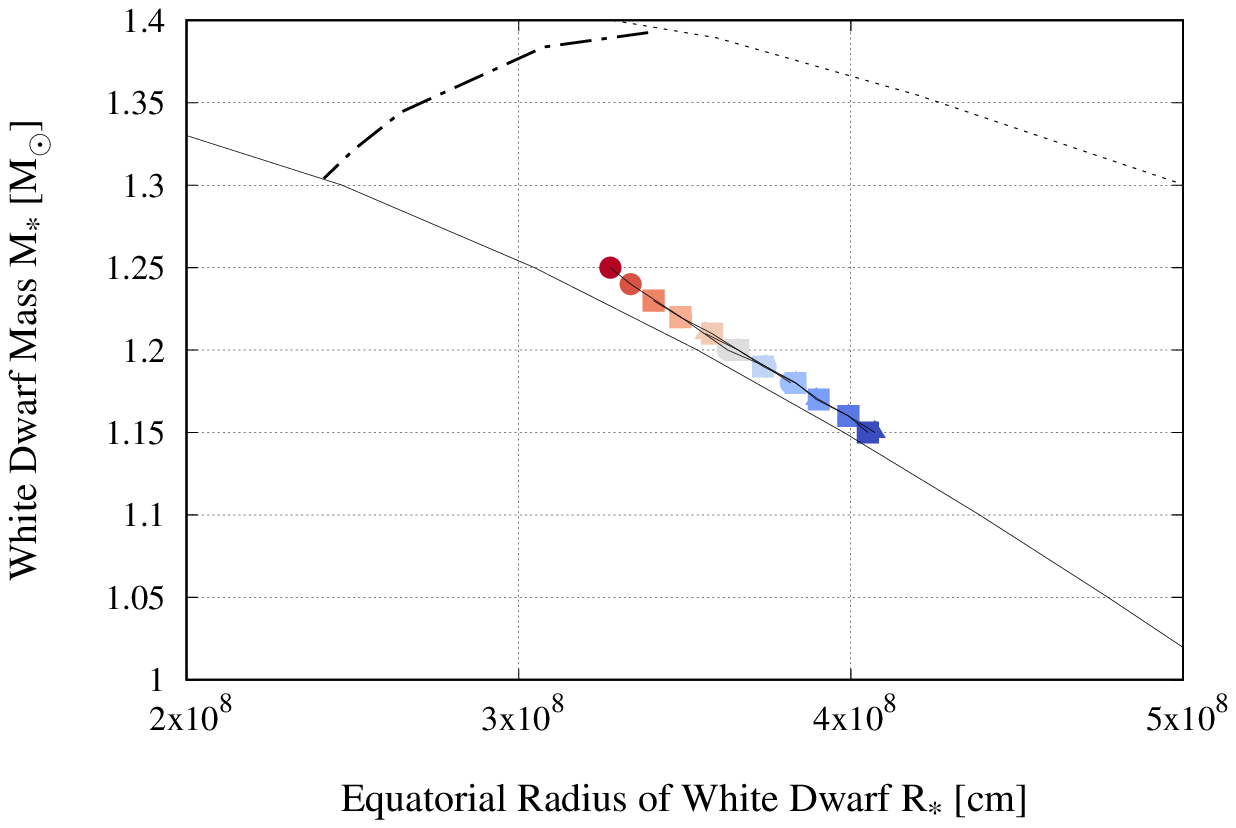}
 \end{minipage}
 \caption{
 White dwarf parameters that give wind solutions compatible with WD J005311. The left and right panels show surface magnetic field and spin angular frequency ($B_*, \Omega$) and mass and radius ($M_*,R_*$), respectively. 
 }\label{fig:j005311wind_summary}
\end{figure}

Let us first describe the basic properties of the wind solutions. 
In Fig. \ref{fig:profile}, we show the case with $M_* = 1.25\,M_\odot$, $R_* = 3.3\times 10^{8}\,\rm cm$, $B_* = 4.2 \times 10^7\, {\rm G}$, $\Omega = 0.5\,\rm s^{-1}$, $v_r(\infty) = 2.4\times 10^9\,{\rm cm\,s^{-1}}$, and $\dot M = 6\times 10^{-6}\,M_\odot\,\rm yr^{-1}$,
that corresponds to the massive end of the allowed parameter region (see Fig. \ref{fig:j005311wind_summary}). 
Other wind solutions in Fig. \ref{fig:j005311wind_summary} basically have same overall features as Fig. \ref{fig:profile}.
The slow, Alfv\'en, and fast points are at $r_{\rm s} = 6.0 \times 10^8\,\rm cm$, $r_{\rm A} = 5.9 \times 10^{9}\,\rm cm$, and $r_{\rm f} = 3.8 \times 10^{12}\,\rm cm$, respectively indicated by the asterisks. 
The photospheric radius of the wind solution is at $r_{\rm ph} \approx r_{\rm A}\sim 10^{10}\,\rm cm$. 
The left-top panel shows the radial and azimuthal velocity profile in linear scale. 
In the left panel of Fig. \ref{fig:comp}, we also plot other velocities including the radial and azimuthal \Alfven velocities, the corotation velocity with the degenerate core, the adiabatic sound velocity of the gas, and the escape velocity.
In the radial direction, the gas is initially subsonic and mainly accelerated by the thermal pressure up to the slow point. 
On the other hand, in the azimuthal direction, the velocity is already trans-sonic at the surface since the gas corotates with the rapidly rotating star. 
In this region, $v_{\phi} \propto r$. 
For $r_{\rm s} < r < r_{\rm A}$, the wind is further accelerated in the radial direction mainly by the magnetic sling effect until the inertia of the wind becomes comparable to the magnetic field. 
After passing the \Alfven point, the radial velocity approaches to the azimuthal \Alfven velocity with the additional acceleration by the pressure gradient of the azimuthal magnetic field and finally passes the fast point. 
The right-top panel of Fig. \ref{fig:profile} shows the magnetic field strength of the wind solution. 
The radial field is simply proportional to $r^{-2}$ due to the constraint equation (Eq. \ref{eq:Br_con}).
Although the constraint equation of the azimuthal field (Eq. \ref{eq:Bphi}) is a bit complicated, 
the solution turns out to be approximately proportional to $r^{-1}$. 
Note that the azimuthal field becomes dominant at slightly beyond the slow point. 

In Fig. \ref{fig:profile}, the density (middle-left), temperature (middle-right), and radiation luminosity (bottom-left) profiles are also shown.
At the surface where carbon burns, the density and temperature are $\rho (R_*) \sim 7 \times 10^4\,\rm g\,cm^{-3}$ and $T(R_*) \sim 7\times 10^8\,\rm K$, respectively.
The radiation luminosity stays almost constant; it slightly decreases with radius for $R_* < r < r_{\rm s}$ where the wind is accelerated by the thermal pressure gradient.
For $r_{\rm s} < r \lesssim r_{\rm ph}$, the gas expands almost adiabatically, i.e., $\rho T^3 \propto r^0$.
Beyond the photosphere, the gas and radiation are decoupled and the temperature becomes constant.  
The right panel of Fig. \ref{fig:comp} shows the profile of the energy density of the wind solution. The black solid line is the effective total energy density (Eq. \ref{eq:e_eff}). We also plot each component in Eq. (\ref{eq:e_eff}); the magnetic field energy, the kinetic energy, the radiation energy, the gas thermal energy, and the gravitational energy. 
The energy density is dominated by the magnetic field energy and stays almost constant even taking into account the radiation loss. 
The kinetic energy grows, i.e., the wind is accelerated, by consuming the thermal energy for $R_* < r < r_{\rm s}$ and the magnetic field energy for $r > r_{\rm s}$. 

Let us now discuss the allowed parameter region of WD J005311. 
Fig. \ref{fig:j005311wind_summary} shows sequences of wind solutions. 
The left panel shows the surface magnetic field and spin angular frequency and the right panel shows the mass and radius. 
The circle, triangle, square symbols represent the cases with $(v_r(\infty), \dot M) = (2.4\times 10^9\,{\rm cm\,s^{-1}},6\times 10^{-6}\,M_\odot\,\rm yr^{-1})$, $(1.9\times 10^9\,{\rm cm\,s^{-1}},6\times 10^{-6}\,M_\odot\,\rm yr^{-1})$, and $(2.4\times 10^9\,{\rm cm\,s^{-1}},3\times 10^{-6}\,M_\odot\,\rm yr^{-1})$, respectively. 
The colors correspond to the y-axis of the right panel; redder (bluer) points correspond to larger (smaller) mass cases. 
In each sequence, a more massive WD has a faster spin.
This can be understood as follows.
In the case of magnetic rotating wind, the radiation luminosity is almost constant with radius, $L_{\rm rad}(R_*) \approx L_{\rm rad}(r_{\rm ph})$ (see the bottom panel of Fig. \ref{fig:profile}).
Thus, the observed bolometric luminosity sets the nuclear burning rate on the surface through Eq. (\ref{eq:inner_boundary}). 
Since the nuclear burning rate is very sensitive to the temperature, it is almost fixed as $T(R_*) \sim 7\times 10^8\,\rm K$, 
and consequently the gas energy density on the surface is almost fixed. 
Since the quasi-hydrostatic equilibrium is achieved in the radial direction around the surface (see Fig. \ref{fig:profile}), the energy density is directly connected to the effective surface gravity $\approx GM_*/R_*^2 - R_*\Omega^2$.  
In general, a more massive WD has a smaller radius, thus a larger gravity.
In order to compete with it, a faster spin is required. 
By combining the $M_*\mbox{-}R_*$ relation with this condition on the effective surface gravity, a relation between $M_*$, $R_*$, and $\Omega$ is almost fixed.
This is shown in the right panel; different sequences align in a line even though they have different $\dot M$ and $v_r(\infty)$.

There exist high and low mass ends in each sequence of solutions in Fig. \ref{fig:j005311wind_summary}. 
The high mass ends are set by Eq. (\ref{eq:rA}); 
from Eqs. (\ref{eq:vA}) and (\ref{eq:v_r_inf}), $r_{\rm A} = ({\cal F}_B^2/{\dot M} v_{\rm A})^{1/2} = [v_r(\infty)^3/v_A \Omega^2]^{1/2}$, thus, for a fixed $v_r(\infty)$, the \Alfven radius becomes smaller for a larger $\Omega$ or larger $M_*$, 
and eventually the condition $r_{\rm A} > (L_{\rm rad, A}/\pi a c T_{\rm A}^4)^{1/2}$ breaks down.
This physically means that a more massive and compact WD satisfying Eqs. (\ref{eq:Mdot}) and (\ref{eq:v_r_inf}) requires to have a radiation luminosity larger than the observed one. 
The upper limit of the mass is larger for a sequence with a larger $v_r(\infty)$. 
We find that $M_* \lesssim 1.3\,M_\odot$ as long as assuming that $v_r(\infty)$ is non-relativistic. 
The low mass end is also set by Eq. (\ref{eq:rA}); a smaller the WD mass becomes, larger $r_{\rm A}$ becomes, and the condition $r_{\rm A} < r_{\rm ph, obs}$ breaks down.
Even if accepting $r_{\rm A} > r_{\rm ph, obs}$ and taking smaller $M_*$, we find that the carbon burning no longer ignites on the surface for $M_* \lesssim 1.1\,M_\odot$. This may be also conflict with the observed chemical abundance of the WD J005311 wind. 
We conclude that the observed properties of WD J005311 can be explained by the rotating magnetic wind from an ONe WD 
with $M_* = 1.1\mbox{-}1.3\,M_\odot$, $B_* = (2\mbox{-}5)\times 10^7\,\rm G$, and $\Omega = 0.2\mbox{-}0.5\,\rm s^{-1}$. 

\section{Discussion}
Contrary to \cite{2019Natur.569..684G}, our model indicates that WD J005311 is not a super-Chandrasekhar-mass system 
and will neither explode as a supernova Ia nor collapse into a neutron star. 
The inferred fast spin and strong magnetic field are still consistent with the WD merger origin as proposed by \cite{2019Natur.569..684G}. 

Although the steady wind solution can only describe the current state of WD J005311, our model has several implications on the fate.
WD J005311 continues to spin down while the wind blows. 
Assuming that $\dot M$ and $B_*$ are constant with time, the spindown luminosity is given by $L_{\rm sd} = \dot M v_r(\infty)^2/2 \propto \Omega^{4/3}$ and the spin angular frequency evolves as $\Omega = \Omega_{\rm i}(1-t/t_{\rm sd})^{3/2}$ where $t_{\rm sd} = 3 I \Omega_{\rm i}^2/(\dot M v_{\infty, \rm i}^2) \sim 2.4\,\rm kyr\,(I/5\times 10^{49}\,{\rm g\,cm^2})(\Omega_{\rm i}/0.5\,\rm s^{-1})^2$. Here $\Omega_{\rm i}$ is the initial spin angular frequency and $I$ is the moment of inertia of the WD.
The spin angular frequency can become practically zero if the wind blows another few kyr. 
On the other hand, the mass in the carbon burning region can be estimated as $\Delta M \approx \rho(R_*)R_*^3 \sim 0.01\,M_\odot$ and the residual lifetime of the wind will be $t_{\rm w} \approx \Delta M/\dot M =$ a few kyr. 
This is consistent with the lifetime of carbon flash obtained by the stellar evolution calculation of the CO WD merger remnant~\citep{2016MNRAS.463.3461S}. 
Since $t_{\rm sd} \sim t_{\rm w}$, the future of WD J005311 is subtle; if $t_{\rm sd} < t_{\rm w}$, it will significantly spin down and join to the observed sequence of massive magnetic WDs, most of which are slowly rotating~\citep{2013ApJ...773...47B}. 
Otherwise, WD J005311 will appear as a fast-spinning strongly magnetized WD.
After the wind driven by nuclear burning blows off, the WD will spin down mainly via the dipole radiation as neutron star pulsars.  
Such a system has been hypothesized as a source of e.g., gamma rays~\citep{1988SvAL...14..258U}, cosmic rays~\citep{2011PhRvD..83b3002K}, and coherent radio bursts~\citep{2016MNRAS.461.1498M}.  

Our model also suggests several strategies for follow up observation of J005311. 
The rotation period of the flow at the photospheric radius is $\approx r_{\rm ph, obs}/v_{\phi, \rm ph} \sim 1\,\rm min$. 
An anisotropy of emission intensity, if any, may induce time variations associated with that. 
Although so far no time variation has been reported, we encourage photometric observations with a good time resolution using e.g., ULTRACAM~\citep{2001NewAR..45...91D} and Tomo-e Gozen~\citep{2016SPIE.9908E..3PS}. 
Deeper spectroscopic observations with a broader wavelength range would be helpful to constrain the wind properties, i.e., the velocity profile, mass loss rate, and chemical composition. 
In principle, the Zeeman splitting can be used to determine the magnetic field strength at the photospheric radius; $B(r_{\rm ph, obs}) \sim 10^5\,\rm G$ in our model, though challenging due to the fact that the emission lines are highly broadened. 
In addition, the infrared nebula may also provide us some information of the WD J005311. 
The rotating magnetic wind may induce the anisotropy in the expansion velocity or the chemical composition of the nebula, that can be investigated by the infrared spectroscopy. 

Our rotating magnetic wind model could be applied also to other nova events. 
For example, MAXI J0158−744 exhibits a very broad emission line of Ne IX \citep{2012ApJ...761...99L, 2013ApJ...779..118M}. Though MAXI J0158−744 event had a peak luminosity exceeding the Eddington limit, the duration might be too short for this radiation to accelerate the wind to the velocity required to reproduce the observed spectrum \citep{2014ApJ...787..165O, 2017symm.conf...49S}.

Let us finally discuss some caveats regarding our wind model. 
First, the convective interface between the wind region and the degenerate core, including the nuclear burning layers, is not solved. 
A global solution combining the stellar evolution of the degenerate core and the rotating magnetic wind is required in particular for predicting the past and the future of WD J005311. 
Second, we use the flux-limit-diffusion approximation. 
In order for calculating the spectrum of the WD J005311 including the emission lines, a more exact radiation transfer needs to be implemented. 
Third, our solution only describes the wind in the equatorial plain. 
The magnetic rotating wind is in general anisotropic. 
A multi-dimensional wind solution is required in order for modeling the observed spectrum and light curve more precisely. 
Those are our future work.

\acknowledgments
We thank Daisuke Nakauchi, Kojiro Kawana, Hajime Kawahara, and Hirotada Okawa for discussion. 
This work is partially supported by JSPS KAKENHI Grant Numbers JP17K14248, JP18H04573, 16H06341, 16K05287, 15H02082, MEXT, Japan.

\software{OPAL code (Iglesias \& Rogers 1996)}

\bibliographystyle{apj}
\bibliography{ref}



\end{document}